\documentclass[aps,prl,twocolumn, superscriptaddress]{revtex4-1}

\usepackage[utf8]{inputenc}
\usepackage{amsmath}
\usepackage{amsfonts}
\usepackage{amssymb}
\usepackage{graphicx}
\usepackage{color}
\usepackage[dvipsnames]{xcolor}
\usepackage{bbm}
\usepackage{rotating}
\usepackage{ulem}
\DeclareMathOperator{\Span}{span}

\begin{document}
\title{Gauge equivariant neural networks for quantum lattice gauge theories}

\author{Di Luo}
\email{diluo2@illinois.edu}
\affiliation{Department of Physics, University of Illinois at Urbana-Champaign, IL 61801, USA}
\affiliation{IQUIST and Institute for Condensed Matter Theory and NCSA Center for Artificial Intelligence Innovation,  University of Illinois at Urbana-Champaign, IL 61801, USA}
\author{Giuseppe Carleo}
\affiliation{Institute of Physics, \'{E}cole Polytechnique F\'{e}d\'{e}rale de Lausanne (EPFL), CH-1015 Lausanne, Switzerland}
\author{Bryan K. Clark}
\affiliation{Department of Physics, University of Illinois at Urbana-Champaign, IL 61801, USA}
\affiliation{IQUIST and Institute for Condensed Matter Theory and NCSA Center for Artificial Intelligence Innovation,  University of Illinois at Urbana-Champaign, IL 61801, USA}
\author{James Stokes}
\email{Corresponding author:    jstokes@flatironinstitute.org}
\affiliation{Center for Computational Quantum Physics and Center for Computational Mathematics, Flatiron Institute, New York, NY 10010 USA}

\date{\today}

\begin{abstract}
Gauge symmetries play a key role in physics appearing in areas such as quantum field theories of the fundamental particles and emergent degrees of freedom in quantum materials. Motivated by the desire to efficiently simulate many-body quantum systems with exact local gauge invariance, gauge equivariant neural-network quantum states are introduced,
which exactly satisfy the local Hilbert space constraints necessary for the description of quantum lattice gauge theory with $\mathbb{Z}_d$ gauge group {and non-abelian Kitaev $D(G)$ models} on different geometries. Focusing on the special case of $\mathbb{Z}_2$ gauge group on a periodically identified square lattice, the equivariant architecture is analytically shown to contain the loop-gas solution as a special case. Gauge equivariant neural-network quantum states are used in combination with variational quantum Monte Carlo to obtain compact descriptions of the ground state wavefunction for the $\mathbb{Z}_2$ theory  away from the exactly solvable limit, and to demonstrate the confining/deconfining phase transition of the Wilson loop order parameter.
\end{abstract}
\maketitle

\paragraph{Introduction --}
Quantum many-body systems defined on a lattice with local gauge invariance occur ubiquitously in the description of physics at diverse energy scales\,---\,they appear in the effective theories of many-electron systems \cite{lee2006doping} and topological phases \cite{kitaev2009topological}, in bosonization of two-dimensional lattice fermions \cite{chen2018exact, bravyi2002fermionic}, and in the microscopically regulated description of interacting elementary particles in the Standard Model \cite{kogut1975hamiltonian}. Quantum lattice gauge theories are characterized by the fact that the physical states span a subspace of the many-body Hilbert space which is defined by satisfying a set of local operator constraints. {These operator constraints gives rise to  group invariance of the associated wavefunction, called gauge invariance.}

Because of the analytic intractability of gauge theories, it is important to develop techniques for simulating them.  
Methods such as lattice-gauge theory \cite{kogut1979introduction} accomplish this via a quantum-classical mapping which rewrites the quantum problems as a statistical mechanics problem in one higher dimension.  This mapping can only be done for sign-free problems without introducing a `negative' weight in the statistical mechanics model which induces either an exponential cost in the simulations or the need for additional approximations to mitigate the sign-problem.  
Alternatively, quantum gauge theories can be simulated using variational methods based on compact parameterizations of many-body wavefunctions. DMRG \cite{PhysRevLett.69.2863}, a variational approach based on matrix-product states, has been extensively used to analyze gauge theories \cite{banuls_review_2020}.  While DMRG is efficacious for one-dimensional systems, the bond-dimension of the matrix product state needed to accurately represent the ground state grows exponentially with width in higher dimensions.  

Machine-learning techniques based on a neural-network variational representation of quantum states \cite{carleo2017solving}, have extended the scope of variational methods by accurately representing low-energy states of strongly correlated systems in two or more spatial dimensions \cite{PhysRevLett.122.226401,PhysRevResearch.2.033429,paulinet,PhysRevResearch.2.023358, PhysRevB.102.205122, PhysRevLett.124.020503,topo_wf,Gao2017,Glasser_2018}. Imposing physical symmetries in neural-network quantum states is a very active research topic \cite{PhysRevLett.121.167204,PhysRevLett.124.097201} that reflects the broader need to impose symmetries in machine learning applications to physics \cite{RevModPhys.91.045002}.
Our work further extends the scope of neural-network quantum states by developing gauge equivariant neural networks which are special-purpose variational families of wave-functions that are explicitly gauge invariant.

Our work is inspired by recent developments of group-invariant networks with group equivariant layers. These have found applications both in data-science and physics. For data-science, the data-generating process is often assumed to be invariant under a symmetry group $G$ \cite{cohen2016group, kondor2018generalization}. For example, Ref.~\onlinecite{cohen2019gauge} derived equivariant layers starting from the assumption that the input data transforms covariantly as a tensor field on a Riemannian manifold. Very recently, gauge equivariant networks have been constructed and variationally optimized using the reverse-relative entropy to approximate the Gibbs distribution associated with the Euclidean action functional for lattice Yang-Mills in the case of abelian $U(1)$ \cite{kanwar2020equivariant} and subsequently non-abelian $SU(N)$ gauge group \cite{boyda2020sampling} (a different gauge-equivariant construction for this non-abelian group is also considered in \cite{favoni2020lattice}). The approach of \cite{kanwar2020equivariant,boyda2020sampling}, which is closely related to variational inference \cite{blei2017variational}, relies on the existence of an analytic continuation between the quantum theory and a positive-definite Gibbs measure, which presents a challenge in the presence of fermions, however, due to the well-known sign problem.

The paper is organized as follows. The $\mathbb{Z}_2$ gauge theory Hamiltonian is introduced using a notation that generalizes to the $\mathbb{Z}_d$ gauge group and higher dimensional models such as 3D toric code and X-cube model (explicitly described in SM III and IV). The general construction of the gauge equivariant neural network is then described. Gauge equivariant neural wavefunctions are variationally optimized using variational Monte Carlo on square lattices up to system size $12 \times 12$, demonstrating the transition of Wilson loop order parameters from perimeter to area law.

\paragraph{ $\mathbb{Z}_2$ Gauge Theory --}
We start by briefly reviewing the formulation of the lattice gauge theory for the simplest non-trivial gauge group $\mathbb{Z}_2 = \{-1,1\}$. The generalization to $\mathbb{Z}_d$ and higher dimensional models, such as 3D toric code and X-cube model, are discussed in Sec. III and IV of the Supplemental Material.  We consider the Hamiltonian with $\mathbb{Z}_2$ fields on the edges $e \in E$ of a  periodic square lattice,
\begin{equation}\label{e:ham}
    H = -J\sum_{f\in F} B_f - h \sum_{e\in E} X_e \enspace ,
\end{equation}
with  $B_f := \prod_{e \in f} Z_e$, where $F$ is the set of the smallest $1 \times 1$ plaquettes $\square$ on the lattice, $e \in f$ are the edges around plaquette $f$ and {$Z_e,X_e$ are the usual Pauli matrices}.  Let $V$ be the set of vertices on the lattice. In order to define the Hilbert space of physical states, define a local operator for each vertex $v \in V$ consisting of the product of Pauli-$X$ operators incident on the given vertex, $A_v := \prod_{e \ni v} X_e$, {where $e \ni v$ indicates the edges containing $v$}. These vertex operators commute amongst themselves, commute with the Hamiltonian $H$ and have eigenvalues $\pm 1$. In this work we focus on the so-called even gauge theory in which the physical Hilbert space $\mathcal{H}_{\rm phys}$ is chosen to be the $+1$ eigenspace of all vertex operators,

\begin{figure}[h]
        \includegraphics[width=1.1\linewidth]{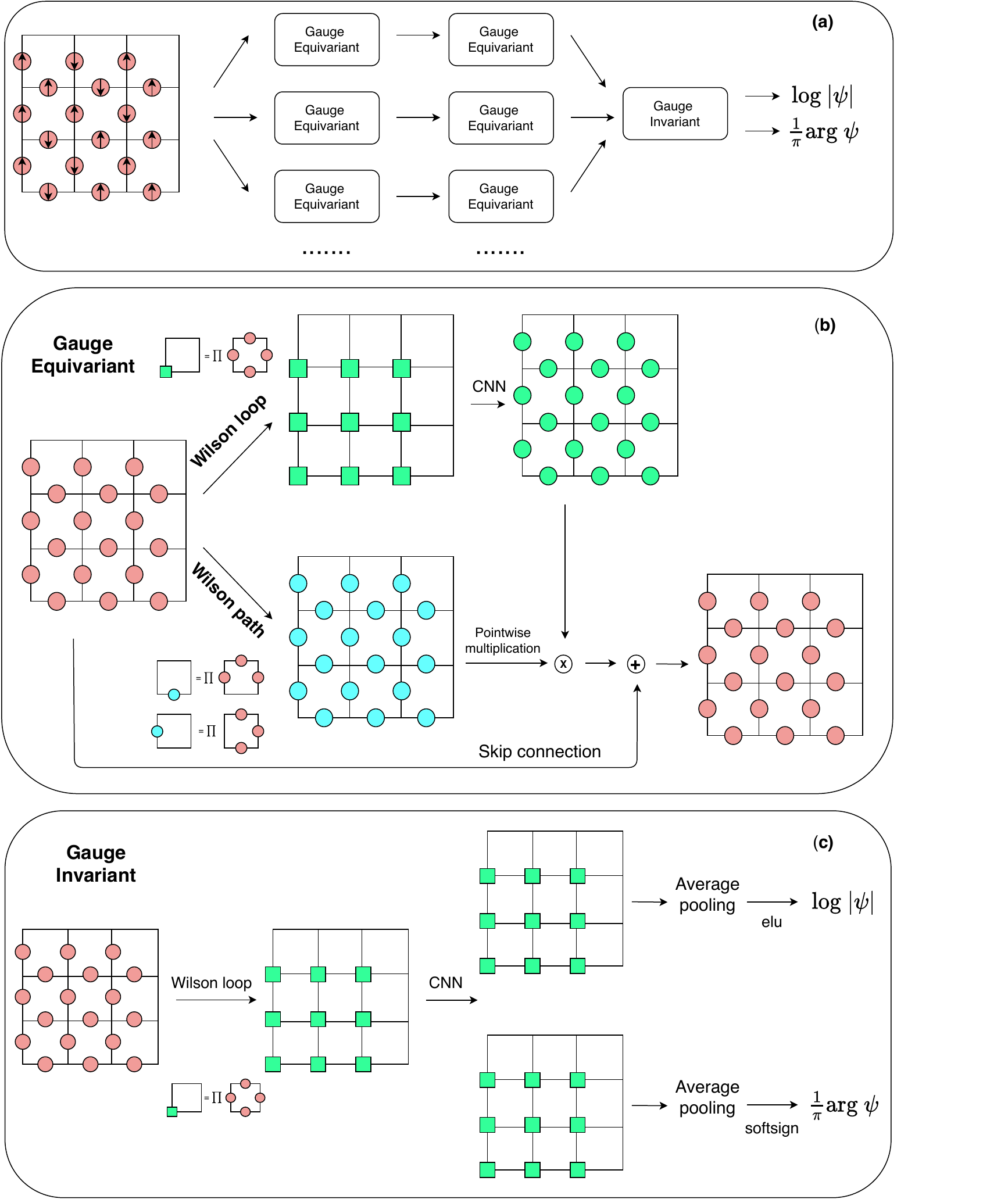}
        \caption{(a) Gauge equivariant neural network architecture.  (b) Gauge equivariant block.  (c) Gauge invariant block. For (b) and (c), the convolution neural network (CNN) component uses channel=2, stride=1, activation function=leaky relu and periodic boundary padding. The kernel size in CNN is different for different systems sizes.}\label{fig:architecture}
\end{figure}

\begin{equation}
    \mathcal{H}_{\rm phys} := \{ |\psi \rangle \in \mathcal{H} : A_v |\psi \rangle = |\psi \rangle \:\: \forall v \in V \} \enspace .
\end{equation}
Since the vertex operators satisfy the global operator identity
$
    \prod_{v \in V} A_v = \mathbbm{1}
$,
it follows that only $|V|-1$ of the constraints defining $\mathcal{H}_{\rm phys}$ are independent. The dimension of physical state space is therefore found to be $ \dim \mathcal{H}_{\rm phys} = \frac{2^{|E|}}{2^{|V|-1}} =  2^{L^2+1}$. Further details on the description of the Hilbert space are provided in Sec. I of the Supplemental Material.

The $\mathbb{Z}_2$ gauge theory is exactly solvable in both the weak coupling ($h \to 0$) and strong coupling ($h\to\infty$) limits. For infinite transverse field $h = \infty$ the non-degenerate ground state is simply the uniform superposition state $|{+} \rangle^{\otimes E}$ which is manifestly gauge invariant, where $|{+}\rangle$ satisfies $X |{+}\rangle = |{+}\rangle$. In the opposite extreme of $h=0$, the ground states are also eigenstates of all $B_f$ and are four-fold degenerate.
As shown originally by Wegner using duality arguments \cite{wegner1971duality}, the uniform superposition $|{+}\rangle^{\otimes E}$ and the ground states at $h=0$ correspond to different phases of matter, which are distinguished by the expectation value of a non-local operator called the Wilson loop 
which is defined for any closed curve $C \subseteq E$ on the lattice as follows
\begin{equation}
    \hat{W}_C := \prod_{e \in C} Z_e \enspace .
\end{equation}
Wegner found a critical value of the transverse field $h_{\rm c}$ separating a deconfined phase for $h < h_{\rm c}$ in which $\langle \hat{W}_C \rangle$ decays exponentially with the perimeter of $C$, from a confined phase where  $\langle \hat{W}_C \rangle$ decays exponentially with the area enclosed by $C$.

\paragraph{Gauge equivariant neural networks --}
We present now a neural network which explicitly preserves the gauge invariance of the wave-function.
The classical configuration space $\mathbb{Z}_2^E$ can be regarded as a subset of the continuous vector space $\mathbb{C}^E := \mathbb{C}^{L \times L \times 2}$ consisting of tensors with shape $(L, L, 2)$, where the edge $e$ is specified by the vertex $v \in V$ indexed by the first two axes and the direction $\mu \in \{\hat{x},\hat{y}\}$ indexed by the third axis.
The components of an arbitrary tensor $\phi \in \mathbb{C}^{E}$ will then be indexed as $\phi_\mu(v)$ where $v \in V$ and  $\mu \in \{\hat{x},\hat{y}\}$.
The action of the gauge group on the space of $(L,L,2)$ tensors is described as follows.
Given a square matrix $\Omega \in \{-1, 1\}^{V} := \{-1, 1\}^{L \times L}$, we define a gauge transformation $g_\Omega : \mathbb{C}^{E} \to \mathbb{C}^{E}$ by the following rule,
\begin{equation}
    (g_\Omega \cdot \phi)_\mu(v) := \Omega(v) \phi_\mu(v) \Omega(v+\mu) \enspace ,
\end{equation}
where $\Omega(v)$ and $\Omega(v + \mu)$ denote the entries of the matrix $\Omega$ at the lattice location $v \in V$ and the shifted lattice location $v + \mu \in V$, assuming boundaries are periodically identified.
It is straightforward to show that the gauge transformation associated with Hilbert space operator $A_v$ is given by $g_{\Omega_v}$ where $\Omega_v$ is defined for each $v' \in V$ by,
\begin{equation}
    \Omega_v(v') =
    \begin{cases}
    -1, & v' = v \\
    +1, & v'\neq v
    \end{cases} \enspace .
\end{equation}
A wave-function which obeys the Gauss law constraint is then one in which 
\begin{equation}\label{eq:invariant_wf}
\Psi(g_{\Omega_v} \cdot \phi) = \Psi(\phi) ,
\end{equation}
for the case where $\phi \in \mathbb{Z}_2^E$. 
Let us call a function $h:  \mathbb{C}^{E} \to \mathbb{C}$ gauge invariant if it satisfies $h(g_\Omega \cdot \phi) = h(\phi)$ for all $g_\Omega$ and a function  $f : \mathbb{C}^{E} \to \mathbb{C}$  gauge equivariant if it satisfies $f(g_\Omega \cdot \phi) = g_\Omega \cdot f(\phi)$ for all $g_\Omega$. A wave-function which consists of multiple layers of gauge equivariant functions interwoven with pointwise nonlinearities followed by a final gauge invariant layer is guaranteed to obey Eq.~\ref{eq:invariant_wf} since group equivariance is preserved by composure and pointwise nonlinearities.  In the following, we construct neural network blocks which are equivariant and invariant respectively.

\paragraph{Gauge Equivariant Layer --}
The construction of the gauge equivariant layer parallels the construction found in \cite{kanwar2020equivariant}, applied to discrete abelian groups. 
Given vertices $v,v' \in V$ and a path $\gamma \subseteq E$ from $v$ to $v'$, consisting of a sequence of adjacent edges, we define the Wilson path as the function $W_\gamma : \mathbb{C}^{E} \to \mathbb{C}$ given by the formula,
\begin{equation}\label{e:wilson_z2}
    W_\gamma(\phi) := \prod_{e \in \gamma} \phi_e \enspace .
\end{equation}

It is easy to show that for any $\Omega$ and any path $\gamma$ from $v$ to $v'$ we have
\begin{equation}\label{e:equivariant}
    W_\gamma(g_\Omega \cdot \phi) = \Omega(v) W_\gamma(\phi) \Omega(v') \enspace .
\end{equation}
The above Wilson path is the fundamental primitive from which the gauge-equivariant layers will be constructed. Note that if $\gamma$ is a closed curve $C$ then the function $W_C$ is gauge invariant.

Consider a fixed equivariant layer described by the equivariant function $f : \mathbb{C}^E \to \mathbb{C}^E$. Each such layer is specified by decorating the edges $e=(v,v+\mu) \in E$ of the lattice with the following data
\begin{itemize}
    \item[1.] A path $\gamma_{e} \subseteq E$ from $v$ to $v+\mu$
    \item[2.] A collection of $n_e \geq 1$ closed curves $C_1^e,\ldots C_{n_e}^e$ 
    \item[3.] A parametrized neural network $h_e : \mathbb{C}^{n_e} \to \mathbb{C}$
\end{itemize}
From the above data we construct a gauge-equivariant function $f: \mathbb{C}^{E} \to \mathbb{C}^{E}$ defined for all $\phi \in \mathbb{C}^E$ by the rule $f : \phi_e \mapsto f_e (\phi)$, where
\begin{equation}\label{e:layer}
    f_e(\phi) := 
    W_{\gamma_e}(\phi)
    h_e\big(W_{C_1^e}(\phi),\ldots,W_{C_{n_e}^e}(\phi)\big) \enspace .
\end{equation}
The equivariance follows directly from Eq.~\eqref{e:equivariant}. The explicit calculation is outlined below for convenience of the reader,
\begin{align}
    f_e(g_\Omega \cdot \phi)
        & = h_e\big(W_{C_1^e}(g_\Omega \cdot \phi),\ldots,W_{C_n^e}(g_\Omega \cdot \phi)\big) W_{\gamma_e}(g_\Omega \cdot \phi) \notag \\
        & = h_e\big(W_{C_1^e}(\phi),\ldots,W_{C_n^e}(\phi)\big) \Omega(v)W_{\gamma_e}(\phi)\Omega(v+\mu) \notag \\
        & = \Omega(v) f_e(\phi) \Omega(v+\mu) \\
        & = \big(g_\Omega \cdot f(\phi)\big)_e \enspace .
\end{align}

\begin{figure}\label{fig:benchmark}
        \includegraphics[width=1\linewidth]{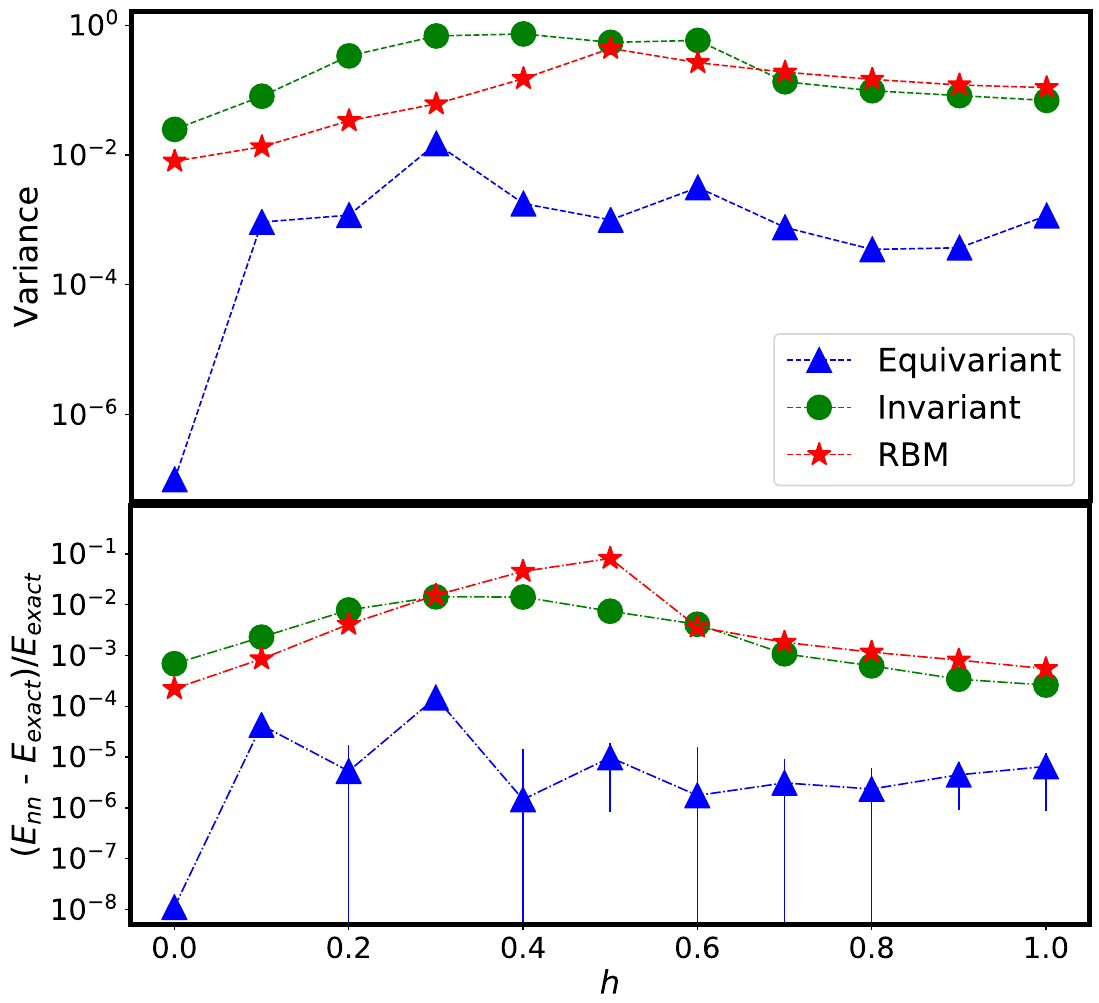}
        \caption{\label{fig:energy3} The  variance {\bf(top)} and the energy difference between the exact ground state {\bf(bottom)} of the gauge equivariant network, gauge invariant network (no equivariant layers) and RBM on a $3 \times 3$ lattice of Eq.~\ref{e:ham}.  The details of the above networks are provided in Sec. VI of the Supplemental Material. }
\end{figure} 

\begin{figure}
        \includegraphics[width=1\linewidth]{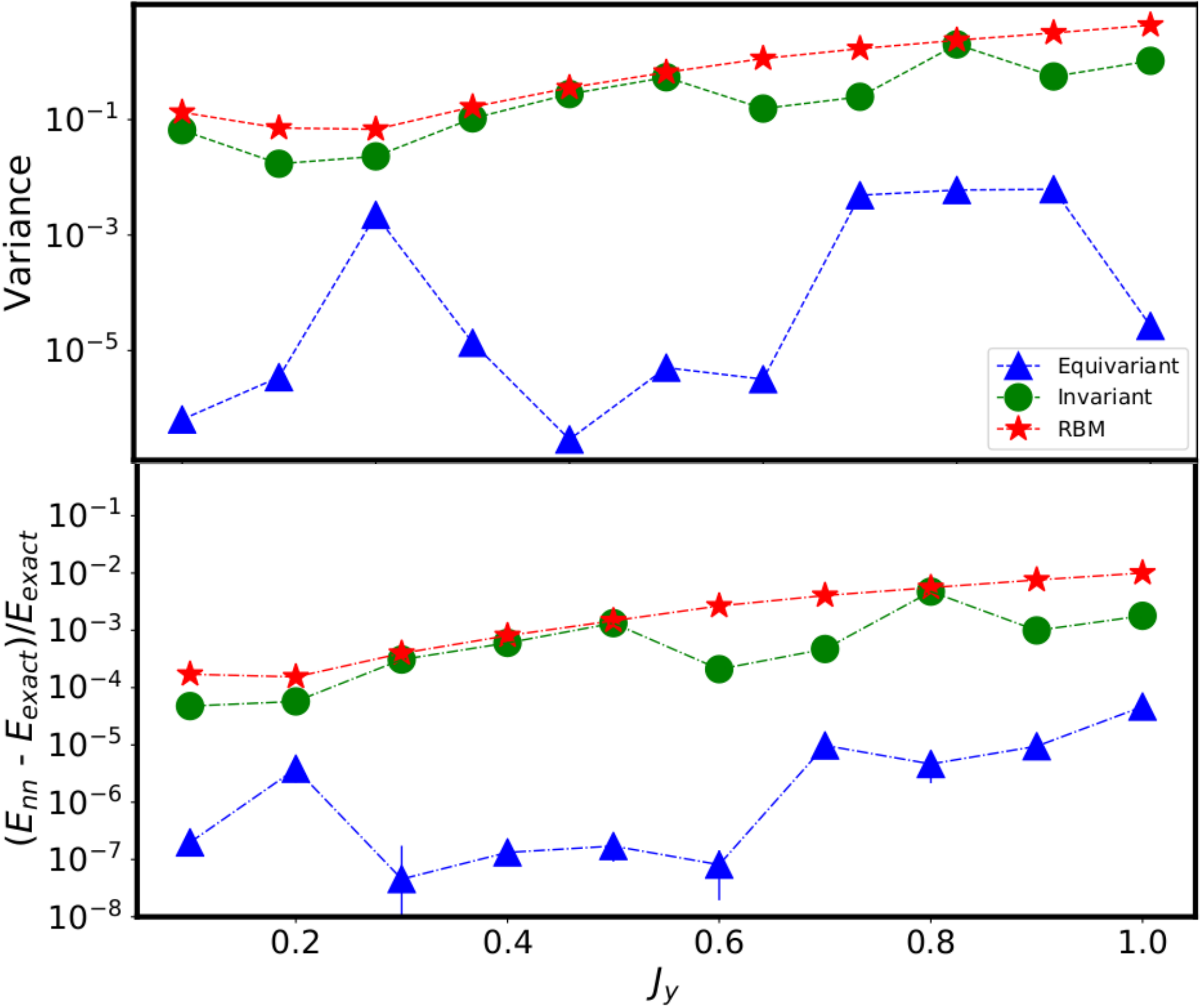}
        \caption{\label{fig:energy_new} {The  variance {\bf(top)} and the energy difference between the exact ground state {\bf(bottom)} of the gauge equivariant network, gauge invariant network (no equivariant layers) and RBM on a $3 \times 3$ lattice of $H = -\sum_{f\in F} \prod_{e \in f} Z_e - \sum_{e\in E} X_e \enspace - J_y \sum_{f\in F} \prod_{e \in f} Y_e$.  The details of the above networks are the same as Fig.~\ref{fig:energy3}.} }
\end{figure}

\begin{figure}
        \includegraphics[width=1\linewidth]{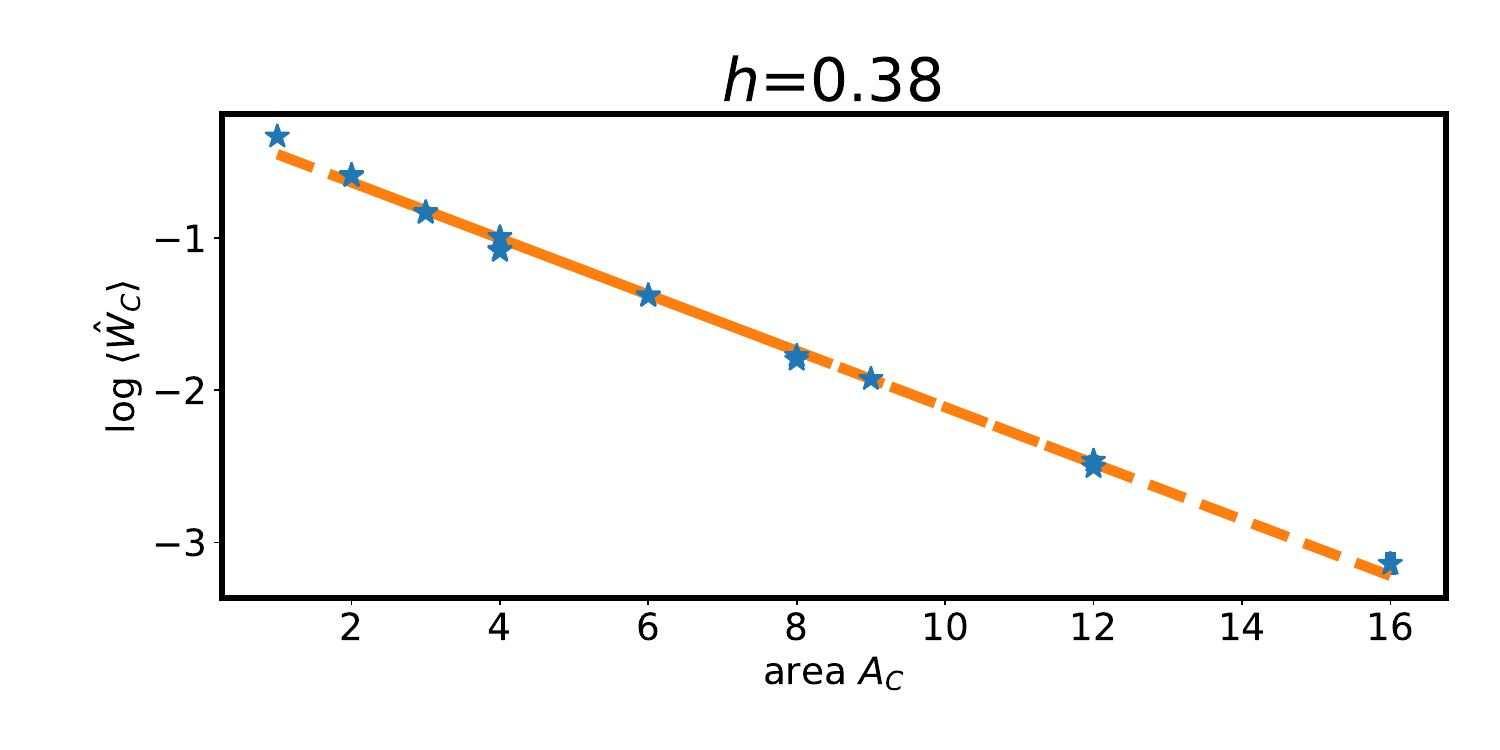}
        \includegraphics[width=1\linewidth]{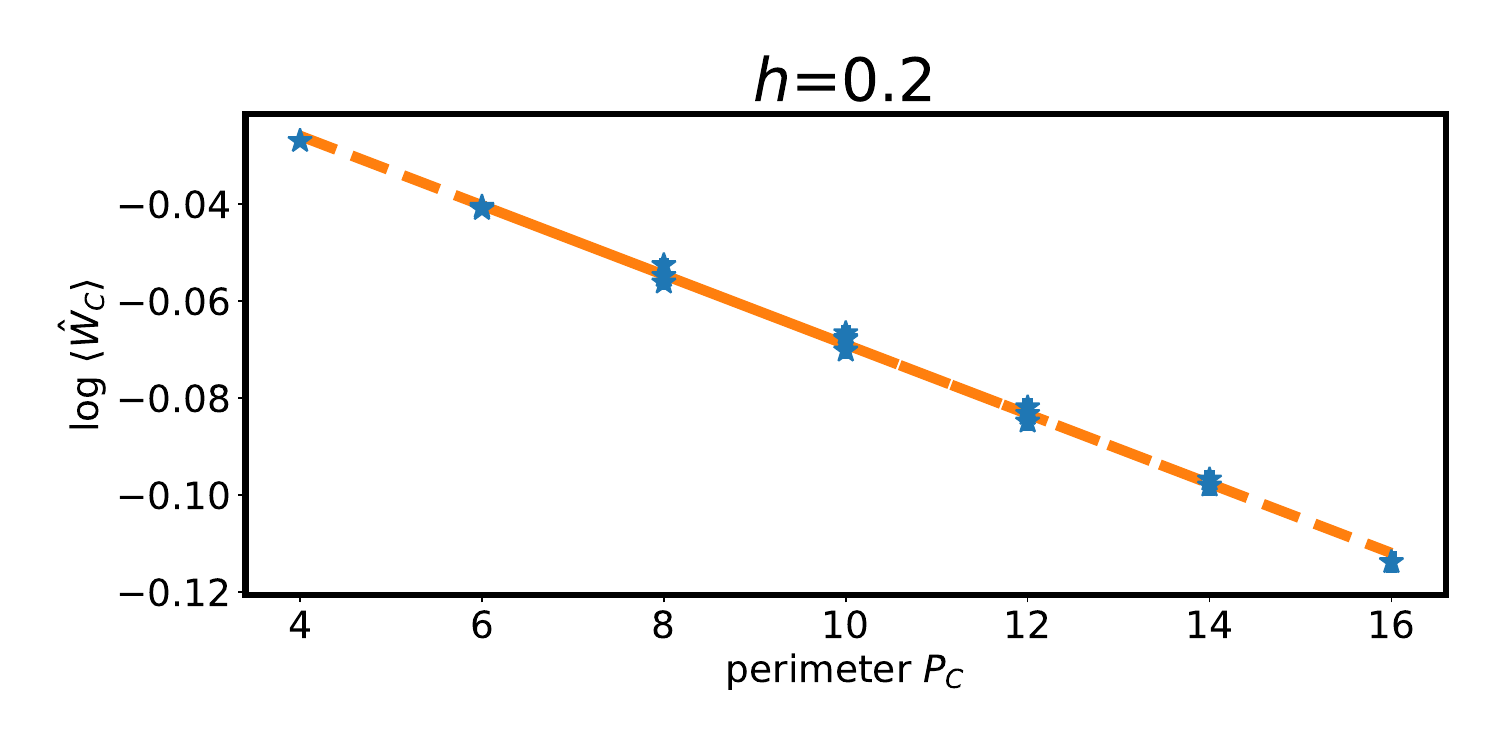}
        \caption{\label{fig:area_law} The ground state expectation value of rectangle Wilson loops of size $l_1 \times l_2$ ($l_1,l_2 \leq 4$) as a function of the enclosed area $A_C$ in the confining phase for $h > h_{\rm c}$ (\textbf{Top}) and of the enclosed perimeter $P_C$ in the deconfined phase for $h < h_{\rm c}$ (\textbf{Bottom}) on a 12 $\times$ 12 lattice. The linear fit to the log-linear plot is consistent with area law scaling $\langle \hat{W}_C \rangle \sim e^{- \alpha A_C}$ with best fit parameter $\alpha = 0.185$ and a perimeter law scaling $\langle \hat{W}_C \rangle \sim e^{- \alpha' P_C}$ with best fit parameter $\alpha' = 0.00718$.\footnote{Due to the smallness of $\alpha'$, we note that a linear fit of $\langle \hat{W}_C \rangle$ versus $P_C$ is also consistent with the data.}}
\end{figure}

\begin{figure}
        \includegraphics[width=1\linewidth]{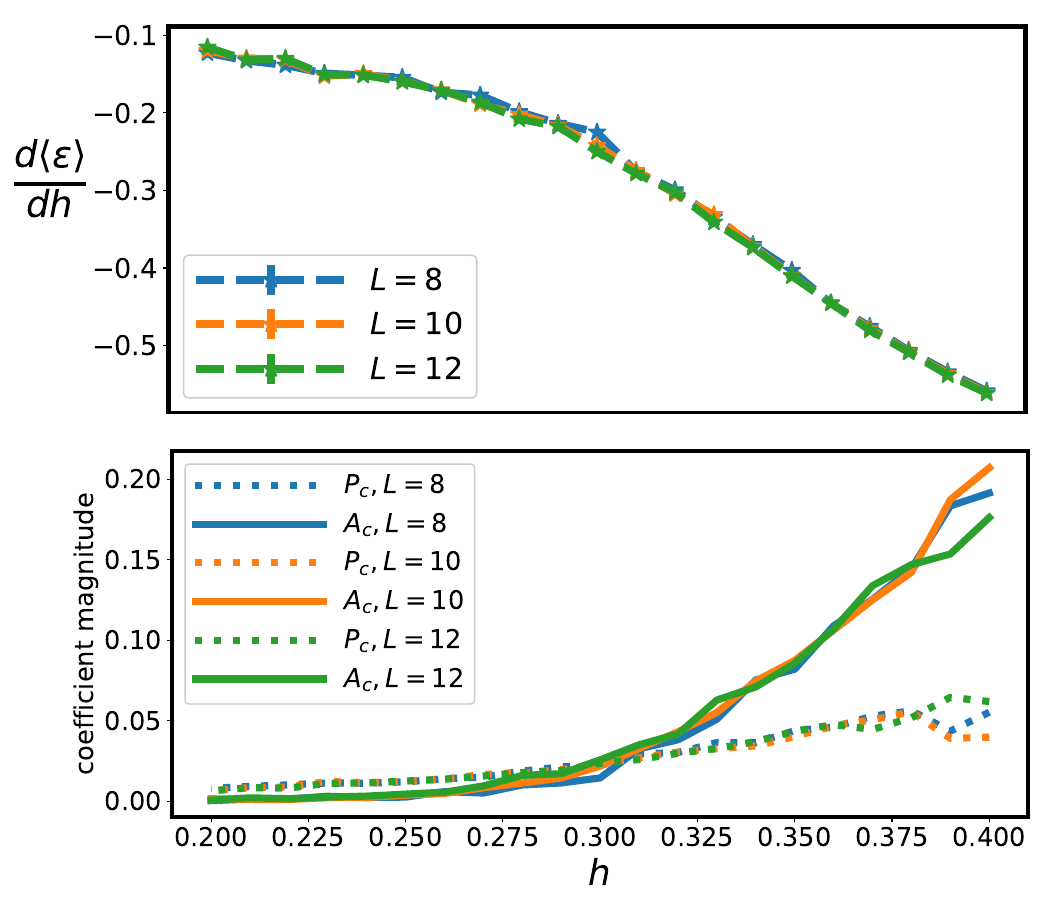}
   \caption{\label{fig:phase_trans} Based on the optimized gauge equivariant states on $L\times L$ lattices as a function of $h$, it shows (\textbf{Top}): derivative of the energy per site calculated by the Hellmann-Feynman theorem. (\textbf{Bottom}): Magnitude of $k_1$ (solid) and  $k_2$ (dotted) for fitting $\log \langle \hat{W}_C \rangle = k_1 A_C + k_2 P_C + b$ with area $A_c$ and perimeter $P_c$. The changes at around $h = 0.30$ in the top figure and the increase in $A_c$ in the bottom figure suggest the confined/deconfined phase transition.}
\end{figure}

\paragraph{Numerical Experiments --}
Here we determine the phase diagram of Eq.~\ref{e:ham} for different values of the transverse field $h$.  When $h=0$, we can analytically write a network which exactly represents the ground state with a single gauge invariant block (and no equivariant blocks), which is detailed in Sec. II of the Supplemental Material. For all $h$, we use variational Monte Carlo to approximately determine the ground states on square lattices.  
The family of variational wavefunctions is summarized graphically in Fig.~\ref{fig:architecture}. It consists of multiple gauge equivariant layers, in which the gauge invariant Wilson loop features within each layer were chosen to consist of all elementary plaquettes of the form $\gamma_e = \square$. The gauge equivariant Wilson path associated with each edge $e$ is chosen to be the curves of the form $\sqcap$ and  $\sqsubset$ ending on $e$. Real-valued weights and biases are used in all neural networks. The neural networks $h_e$ are chosen to be convolutional neural network with periodic padding to capture symmetry and facilitate transfer learning. Residual layers or equivalently skip connections, which are manifestly gauge equivariant, are also employed. The neural network parameters are optimized using the Stochastic Reconfiguration algorithm \cite{Sorella2017book}. Further details about the architecture and optimization are provided in Sec. V of the Supplemental Material.

We start with a benchmark of the hamiltonian in Eq.~\ref{e:ham} on a $3 \times 3$ square lattice by comparing gauge equivariant neural network, gauge invariant network, restricted Boltzmann machine (RBM) and exact diagonalization, where Fig.~\ref{fig:energy3} presents the energy and variance. {In addition, we benchmark a hamiltonian with a sign problem $H = -J\sum_{f\in F} \prod_{e \in f} Z_e - h \sum_{e\in E} X_e \enspace - J_y \sum_{f\in F} \prod_{e \in f} Y_e $ for $J=h=1$, which results are shown in Fig.~\ref{fig:energy_new}}. The number of variational parameters for the above three neural networks are 66, 24 and 1044 respectively. It can be seen that even with small number of parameters, the gauge equivariant neural network achieves better performance than the RBM and attains accurate results close to the exact. We further apply our method to larger square lattices of size $L \times L$ with $L \in \{8,10,12\}$. It is known that the Wilson loop expectation value $\langle \hat{W}_C \rangle$ decays exponentially with area law for $h > h_c$ and with perimeter law for $h < h_c$ \cite{Gregor_2011}. In Fig.~\ref{fig:area_law}, the variational wave function is shown to capture the area law and the perimeter law behaviors of the Wilson loop in the corresponding regimes and attains the related decay factors. In Fig.~\ref{fig:phase_trans}, we compute the energy derivatives and perform a simultaneous fitting of area and perimeter law for log$\langle \hat{W}_C \rangle$ with different $h$. The changes of the data at around $h=0.3$ suggest a deconfinement/confinement phase transition, which is consistent with \cite{vidal2009low, wu2012phase}.

\paragraph{Discussion and future directions --}
In this work, we have showed how to use gauge equivariant networks to represent variational states which exactly obey local gauge constraints.  This significantly expands the space of models whose phase diagrams can now be numerically established.  For example one could variationally explore the phase diagrams of the $\mathbb{Z}_d$ lattice gauge theory for $d>2$ (see SM III); the 3D Toric-Code (SM IV); X-cube fracton model (SM IV) {or Kitave $D(G)$ models (SM V)} with external field; or models with disorder.  Another interesting application would be to relax the restriction to the even sector of the gauge theory and explore the physics of different gauge sectors.  Beyond the explicit constructions given here, it will be interesting to further extend the reach of such networks by generalizing the approach described in this work to models with different gauge symmetries or constraints from gauging subsystem symmetries. Finally, the ability to exactly employ gauge constraints variationally has potential to have impact beyond ground state calculations for example in overcoming obstacles in optimizing combinatorial structures \cite{shalev2017failures} or in successful quantum-state tomography \cite{2018NatPh..14..447T}.

\quad

\textit{Acknowledgements.} Di Luo acknowledges helpful discussion with Jiayu Shen, Luke Yeo, Oleg Dubinkin, Ryan Levy, Peijun Xiao and Ruoyu Sun. We acknowledge support from the Department
of Energy grant DOE de-sc0020165.  The work is partially done with intern and computational support from the Center for Computational Mathematics at the Flatiron Institute. The numerical experiments were conducted using NetKet~\cite{netket:2019}.

\bibliography{references.bib}

\clearpage
\onecolumngrid

\renewcommand\thefigure{S\arabic{figure}}  
\renewcommand\thetable{S\arabic{table}}  
\renewcommand{\theequation}{S\arabic{equation}}
\renewcommand{\thepage}{P\arabic{page}} 
\setcounter{page}{1}
\setcounter{figure}{0}  
\setcounter{table}{0}
\setcounter{equation}{0}

\appendix
\section{\label{app:lindblad} \large{Supplemental Material}}

\section{I. Detailed Description of the $\mathbb{Z}_2$ Gauge Theory Hilbert Space}

The lattice underlying the gauge theory is chosen to be the $L \times L$ square lattice with periodic boundary conditions and topology of the torus. Let $G=(V,E)$ denote the undirected interaction graph with $|V|=L^2$ vertices and $|E|=2L^2$ edges and $|F|=L^2$ faces, in accordance with Euler's formula $|V|-|E|+|F| = 2-2g$ for a torus, which has genus $g=1$.  Vertices, edges and faces are indexed by $v \in V$, $e \in E$, and $f \in F$ respectively. Each edge hosts a qubit $\mathbb{C}^2 = \Span_\mathbb{C} \{|{-1}\rangle , |1\rangle \}$ with basis vectors labeled by the eigenvalues\footnote{Please note that this convention differs from the quantum information literature.} of the Pauli-$Z$ operator $Z|{\pm1}\rangle = \pm |{\pm1}\rangle$, so the joint tensor product Hilbert space for all qubits is $\mathcal{H} = \bigotimes_{e \in E}\mathbb{C}^2$ with orthonormal basis elements $|x\rangle : = \bigotimes_{e \in E}|x_e\rangle$ where $x_e \in \mathbb{Z}_2$. It will be convenient to define the superposition state $|{+}\rangle = \frac{1}{\sqrt{2}}(|{-1}\rangle + |1\rangle) \in \mathbb{C}^2$. Cardinalities of sets are omitted when they appear in superscripts so, for example, $x \in \mathbb{Z}_2^{E}$ denotes a $\pm 1$ string of length $|E|$, indexed by the edges of the graph $G$.

\section{II. Exact Representation of $\mathbb{Z}_2$ Toric Code Ground States}

{ In this section, we provide exact constructions of gauge equivariant neural networks for the ground states and the excited state of the $\mathbb{Z}_2$ toric code model.  This can be done with a neural network with no equivariant blocks and a single invariant block.  It differs from our construction in the main text in two additional ways: (1) we use a fully connected network instead of a convolution network and (2) in addition to all the primitive plaquettes we include two topologically non-trivial loops around the torus.  

There are four degenerate ground states of the $\mathbb{Z}_2$ model. One exact ground state can be obtained by starting with the uniform superposition state $|{+}\rangle^{\otimes E}$ and applying the following projection operator which commutes with all vertex operators,
\begin{equation}
    P := \prod_{f \in F}\frac{\mathbbm{1}+B_f}{2} \enspace .
\end{equation}
which gives rise to the loop-gas ground state $|\psi_{\textrm{loop-gas}}\rangle = P |{+}\rangle^{\otimes E}$. Excited states are obtained from the ground state by relaxing eigenvalues of various plaquette operators $B_f$ to $-1$, in a manner consistent with the global operator constraint $\prod_{f \in F} B_f = \mathbbm{1}$.  Because the toric code is frustration free,  to be a ground state of Eq.~\ref{e:ham}, it suffices to be a ground state of $B_f$ and $X_e$ separately. As the gauge equivariant construction already ensures the system is a ground state of $X_e$ we simply choose a function of the elementary Wilson loops $W_f(\phi)$ for each plaquette $f \in F$

\begin{equation}\label{e:ground_state}
    f(\phi) = 
    h_e\big(W_{C_1}(\phi),\ldots,W_{C_{L^2}}(\phi),W_{C_{x}}(\phi),W_{C_{y}}(\phi) \big)
\end{equation}
for our network
where each $C_{i}$ is the elementary plaquette $\square$ for $i=1,...,L^2$, $C_{x}$ is a loop across the whole torus from x-direction and  $C_{y}$ is a loop across the whole torus from y-direction. 

We must choose a $h_e$ such that Eq.~\ref{e:ground_state} is a ground state for each $B_f$, i.e. $\langle x | B_f |\psi\rangle = \langle x |\psi\rangle$ for each $B_f$ and $x$. It implies that $\langle x | \psi\rangle = 0$ for any $|x \rangle$ such that $B_f |x \rangle = - |x\rangle$. Consider two operators $\tau_z^x = \prod_{i \in C_{x}^e} Z_i$ and $\tau_z^y = \prod_{i \in C_{y}^e} Z_i$. $\tau_z^x$ and $\tau_z^y$ further distinguishes the four degenerate ground state, where the expectation values $a,b$ of the two operators are from one of the choices in $\{\pm 1, \pm 1\}$. Hence, we define $h_e$ as follows
\begin{equation}
    h_e\big(W_{C_1}(\phi),\ldots,W_{C_{n}}(\phi),W_{C_{x}}(\phi),W_{C_{y}}(\phi) \big)= \prod_i h_1(W_{C_i}(\phi)) h_x(W_{C_x}(\phi)) h_y(W_{C_y}(\phi))
\end{equation}
where $h_1(W_{C_i}(\phi))=1$ for $W_{C_i}(\phi)=1$ and zero otherwise, $h_x(W_{C_x}(\phi))=1$ for $W_{C_x}(\phi)=a$ and zero otherwise, $h_y(W_{C_y}(\phi))=1$ for $W_{C_y}(\phi)=b$ and zero otherwise. The above construction provides an exact representation of a one-layer gauge invariant network for each of the ground state of $\mathbb{Z}_2$ toric code. }

\section{III. Gauge Equivariant Construction for $\mathbb{Z}_d$ Toric Code}

\begin{figure}[h]
        \includegraphics[width=0.55\linewidth]{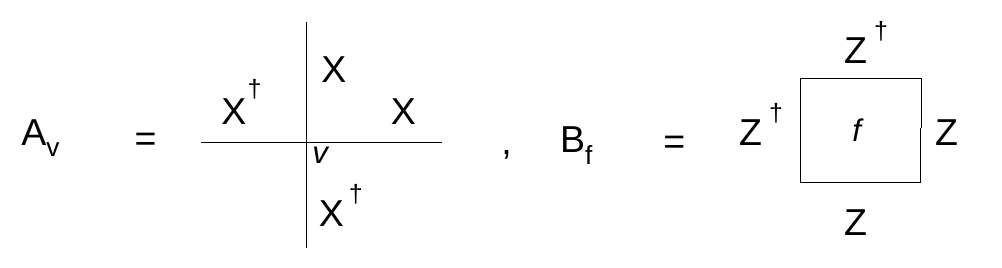}
   \caption{\label{fig:operator_Zd} $A_v$ and $B_f$ in the $\mathbb{Z}_d$ model.}
\end{figure}

In this section we briefly review the generalization to the $\mathbb{Z}_d$ gauge group \cite{Kitaev_2003}. Consider the usual $L \times L$ square periodic lattice where each edge hosts a $d$-dimensional qudit. The Hamiltonian defined on $\mathcal{H} (\mathbb{C}^d)^{\otimes E}$ is
\begin{equation}
    H = -\sum_{v \in V} \sum_{h \in \mathbb{Z}_d} (A_v)^h - \sum_{f \in F} \sum_{h \in \mathbb{Z}_d} (B_f)^h
\end{equation}
where $A_v$ and $B_f$ are shown in Fig.~\ref{fig:operator_Zd} with
\begin{equation}
    X = \sum_{h \in \mathbb{Z}_d} |h+1\rangle \langle h | \enspace , \quad \quad Z = \sum_{h \in \mathbb{Z}_d} \omega^h |h\rangle \langle h | \enspace, \quad \quad \omega = e^{i2\pi /d}
\end{equation}

To discuss the gauge equivariant and invariant construction for the $\mathbb{Z}_d$, we represent the basis for qudit on each edge as $\{e^{i2k\pi/d}|k=0,1,...,d-1\}$. The action of the gauge group on the space of $(L,L,2)$ tensors is described as follows.
Given a square matrix $\Omega \in \{e^{i2k\pi/d}|k=0,1,...,d-1\}^{V} := \{e^{i2k\pi/d}|k=0,1,...,d-1\}^{L \times L}$, we define a gauge transformation $g_\Omega : \mathbb{C}^{E} \to \mathbb{C}^{E}$ by the following rule,
\begin{equation}
    (g_\Omega \cdot \phi)_\mu(v) := \Omega(v) \phi_\mu(v) \Omega(v+\mu)^* \enspace ,
\end{equation}
where $\Omega(v)$ and $\Omega(v + \mu)$ denote the entries of the matrix $\Omega$ at the lattice location $v \in V$ and the shifted lattice location $v + \mu \in V$, assuming boundaries are periodically identified.

It is straightforward to show that the gauge transformation associated with Hilbert space operator $(A_v)^h$ is given by $g_{\Omega^h_v}$ where $\Omega^h_v$ is defined for each $v' \in V$ by,
\begin{equation}
    \Omega^h_v(v') =
    \begin{cases}
    e^{i2\pi h/d}, & v' = v \\
    1, & v'\neq v
    \end{cases} \enspace .
\end{equation}

Given vertices $v,v' \in V$ and a directional path $\gamma \subseteq E$ from $v$ to $v'$, consisting of a sequence of adjacent edges with arrow direction, we define the Wilson path as the function $W_\gamma : \mathbb{C}^{E} \to \mathbb{C}$ given by the formula,
\begin{equation}\label{e:wilson_zd}
    W_\gamma(\phi) := \overrightarrow{\prod_{e \in \gamma}} \phi_e \enspace .
\end{equation}
where the product is directional such that it takes original value $\phi$ when the path $\gamma$ direction on the edge goes upward and right and conjugate value $\phi^*$ when the path $\gamma$ direction on the edge goes down and left.

An exact ground state can be found starting with the uniform superposition state $\big(\frac{1}{\sqrt{d}}\sum_{h\in\mathbb{Z}_d} |h\rangle\big)^{\otimes E}$ and applying the projection operator $P = \prod_{f \in F} \frac{1}{d}\sum_{h \in \mathbb{Z}_d} (B_f)^h$.
It can be shown that the amplitudes in the standard basis are of this state are computed by the following single-layer gauge-invariant neural network,
\begin{equation}
    f(\phi) = \prod_{f \in F} \sum_{h \in \mathbb{Z}_d} W_f(\phi)^h \enspace .
\end{equation}
The invariant and equivariant features consist of Wilson paths corresponding to closed loops and open paths ending on a given edge. The equivariance follows from the identity $W_\gamma(g_\Omega \cdot \phi) = \Omega(v) W_\gamma(\phi) \Omega(v')^*$.

\section{IV. Gauge Equivariant Construction for 3D Toric Code and X-cube Fracton Model}

In this section, we generalize the gauge equivariant neural network construction to the 3D toric code and X-cube fracton model. The Hamiltonian of 3D toric code is given by,
\begin{equation}
    H = -\sum_{v \in V} A_v- \sum_{f \in F} B_f
\end{equation}
where $A_v := \prod_{e \ni v} X_e$ is the product of Pauli-$X$ over the six edges incident on a vertex $v$ and $B_f := \prod_{e \in f} Z_e$ is the product of Pauli-$Z$ over each square face $\square$ of the lattice. The Gauss law constraint can be imposed using a gauge-equivariant wavefunction in which the invariant and equivariant features are chosen to be Wilson paths corresponding to closed loops and open paths ending on a given edge. The Wilson path function is the same form given in Eq.~\ref{e:wilson_z2} in the main text.

\begin{figure}[h]
        \includegraphics[width=0.55\linewidth]{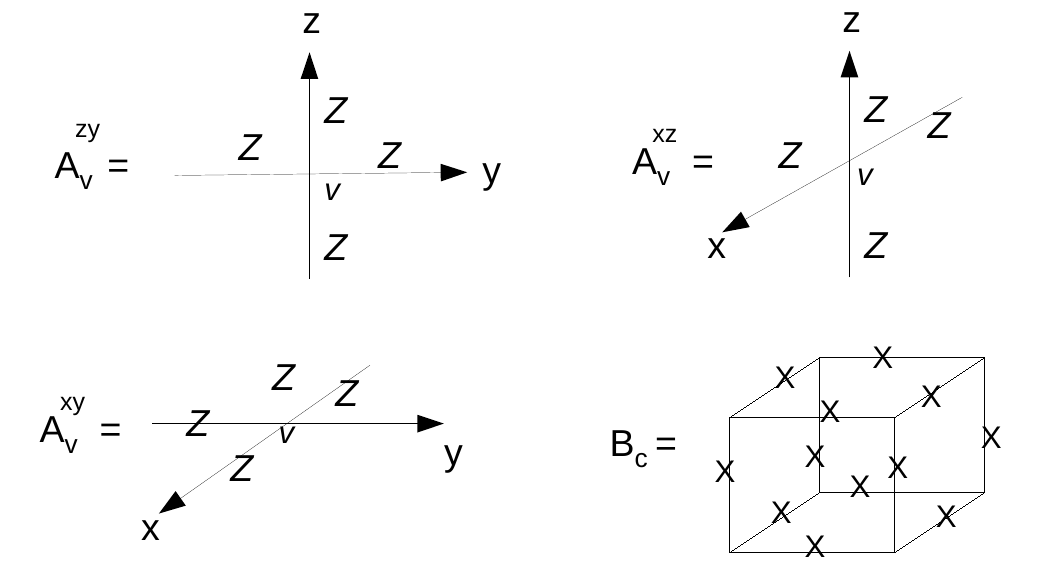}
   \caption{\label{fig:fracton2} $A_v^i$ and $B_c$ in X-cube model.}
\end{figure}

\begin{figure}[h]
        \includegraphics[width=0.55\linewidth]{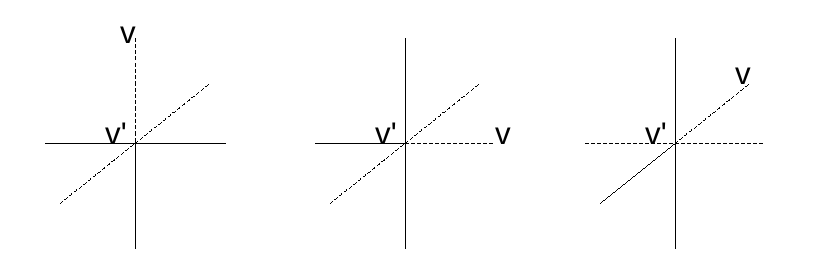}
   \caption{\label{fig:wilson_fracton2} Equivariant features on an edge $v'v$ in the X-cube model can be constructed from curves described by solid lines shown.}
\end{figure}

The  X-cube fracton model is defined on the same cubic lattice and the Hamiltonian is given by
\begin{equation}
    H = -\sum_{v \in V, i} A_v^i- \sum_{c \in C} B_c
\end{equation}
where each of the terms are defined in Fig.~\ref{fig:fracton2}. The Gauss law constraint can again be imposed using a gauge-equivariant wavefunction. Invariant features can be constructed from Wilson paths evaluated on the star-shaped curves shown in Fig.~\ref{fig:fracton2}. Equivariant features on an edge $v'v$ can be constructed from Wilson paths evaluated on the  curves shown in Fig.~\ref{fig:wilson_fracton2}.

The ground state wavefunction amplitudes of 3D toric code can be represented exactly by a gauge invariant neural network of the form \eqref{e:ground_state} where the closed Wilson loops correspond to each square face $\square$ of the cubes. The ground state wavefunction amplitudes of the X-cube model can likewise be represented as a product of Heaviside step functions of Wilson paths, where now the Wilson paths correspond to each star-shape curves in x-y, y-z and x-z direction of the cube as shown in Fig.~\ref{fig:fracton2}.

\section{V. Gauge Equivariant Construction for Kitaev $D(G)$ Models}

{
In this section we provide gauge equivariant neural network construction for general Kitaev Models over $\mathbb{Z}_d$ group and non-abelian groups. Consider the usual $L \times L$ square periodic lattice where each edge has a basis $\{|g\rangle, g \in G\}$ for certain group $G$. We focus on finite group here and group $G = \mathbb{Z}_d$ for $\mathbb{Z}_d$ theory. Without loss of generality, we attach an upward arrow for each edge in y-direction a right arrow for each edge in x-direction. We introduce operators $A_v^g$ and $B_f$ as follows

\begin{figure}[h]
        \includegraphics[width=1\linewidth]{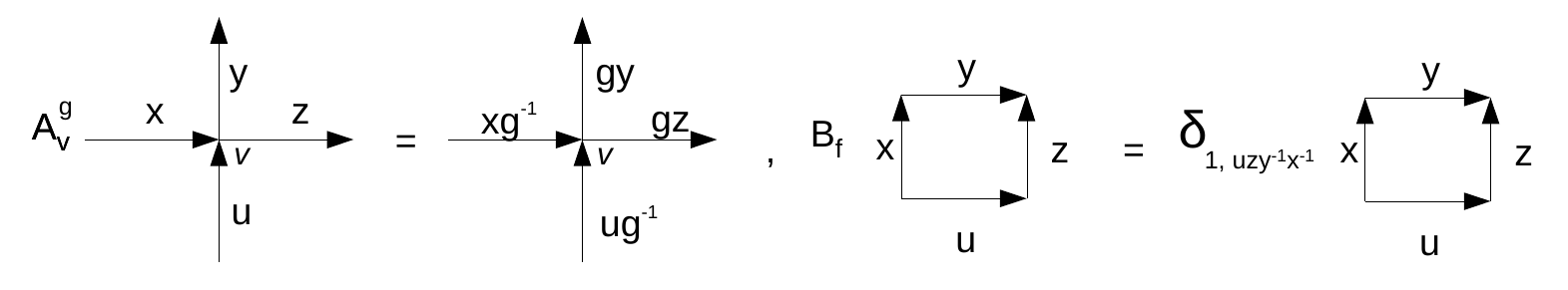}
   \caption{\label{fig:operator} $A_v^g$ and $B_f$ operators.}
\end{figure} 

The Hamiltonian defined on $\mathcal{H} (G)^{\otimes E}$ is 
\begin{equation}
    H = -\sum_{v \in V} A_v- \sum_{f \in F} B_f
\end{equation}
where $A_v = \frac{1}{|G|} \sum_{g \in G} A_v^g$ is the Gauss' law and the gauge constraint.

Define $|{+}\rangle = \frac{1}{\sqrt{|G|}} \sum_{g \in G} |g\rangle$, then the ground state $|\psi\rangle = \prod_{f \in F} B_f |{+}\rangle^{\otimes E}$. This is because $|\psi \rangle$ is both ground state for each $A_v$ and $B_f$. It is easy to verify that $B_f |\psi\rangle = |\psi\rangle$. To see $A_v |\psi\rangle =|\psi\rangle$, notice that $A_v$ and $B_f$ commute with each other and $A_v |{+}\rangle^{\otimes E} =|{+}\rangle^{\otimes E}$.  

Given a configuration $x$ on the lattice, vertices $v_1,v_2 \in V$ and a directional path $\gamma \subseteq E$ from $v_1$ to $v_2$, consisting of a sequence of adjacent edges with direction, we define the generalized Wilson path as the function $W_\gamma : G^{E} \to G^{E}$ given by the formula,
\begin{equation}
    W_\gamma(\phi) :=\overrightarrow{\prod_{e \in \gamma}} \phi_e \enspace .
\end{equation}
where the product is directional such that it takes original value $g$ on the edge when the path $\gamma$ direction agrees with the edge direction and inverse value $g^{-1}$ when the path $\gamma$ direction is opposite to the edge direction. 

We claim that for any open path $\gamma$, $W_{\gamma}$ is gauge equivariant, i.e. $W_{\gamma}$ commutes with $A_v^g$ for each $v$ and $g$. To see this, we consider three cases. The first case is that $A_v^g$ does not act on any edge in the path $\gamma$ and it is clear that they commute. The second case is that $A_v^g$ acts on a vertex along the path $\gamma$ but not $v_1, v_2$. Notice that $A_v^g$ must touch two adjacent edges at the same time and due to the arrow convention, the effect of $A_v^g$ will also cancel out in the product. The last case is that $A_v^g$ is on one of the vertex $v_1, v_2$ of the path $\gamma$, it is straightforward to verify directly $A_v^g$ commutes with $W_\gamma$. Without loss of generality, we consider the a Wilson path function on the bottom edge of a plaquette with path $\gamma$ going clockwise. One can check the following

\begin{figure}[h]
        \includegraphics[width=0.55\linewidth]{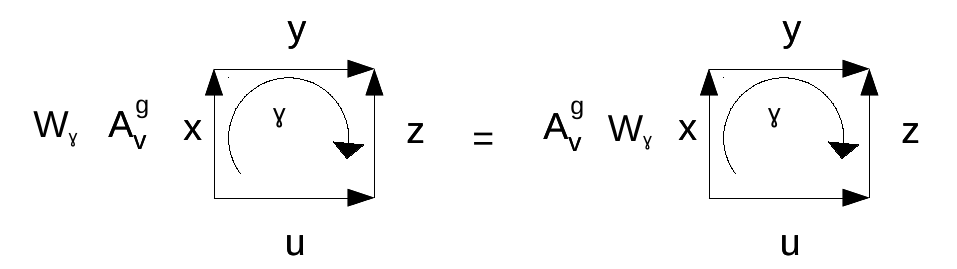}
   \caption{\label{fig:operator} $W_{\gamma}$ commutes with $A_v^g$.}
\end{figure}

For a closed loop $C$, we also introduce a Wilson loop function $W_{C} : G^{E} \to \mathbb{C}$
\begin{equation}
    W_C(\phi) := \text{tr} \overrightarrow{\prod_{e \in \gamma}} \phi_e \enspace .
\end{equation}
Notice that $W_{C}$ is gauge invariant, i.e. $W_{C} \circ A_v^g(\phi) = W_{C}(\phi)$ for any $v$ and $g$. This is because any $A_v^g$ touches either no edge or two adjacent edges along $C$. The invariance property holds clearly if no edge is touched and still holds if two edges are touched since the effect of $A_v^g$ always cancel out due to the arrow convention. Without loss of generality, this can be verified on each plaquette as follows

\begin{figure}[h]
        \includegraphics[width=0.65\linewidth]{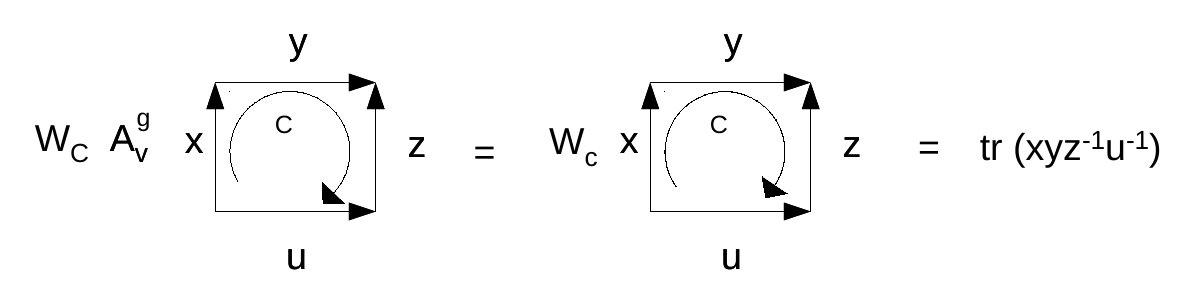}
   \caption{\label{fig:operator} $W_{C}$ is gauge invariant.}
\end{figure}

With $W_\gamma$ and $W_C$, we can construct gauge equivariant and invariant layer for the gauge equivariant neural network with Eq.~\ref{eq:equ_layer} and Eq.~\ref{eq:inv_layer} for general Kitave model with both $\mathbb{Z}_d$ and non-abelian group $G$. A gauge-equivariant function $f: \mathbb{C}^{E} \to \mathbb{C}^{E}$ is defined for all $\phi \in \mathbb{C}^E$ by the rule $f : \phi_e \mapsto f_e (\phi)$, where
\begin{equation}\label{eq:equ_layer}
    f_e(\phi) := 
    W_{\gamma_e}(\phi)
    h_e\big(W_{C_1^e}(\phi),\ldots,W_{C_{n_e}^e}(\phi)\big) \enspace .
\end{equation}
and a gauge-invariant function $h: \mathbb{C}^{E} \to \mathbb{C}$ is defined for all $\phi \in \mathbb{C}^E$ by the rule $h : \phi_e \mapsto h_e (\phi)$, where
\begin{equation}\label{eq:inv_layer}
    h_e(\phi) := 
    h_e\big(W_{C_1^e}(\phi),\ldots,W_{C_{n_e}^e}(\phi)\big) \enspace .
\end{equation}

In particular, we can construct a one-layer gauge invariant network for the ground state as follows
\begin{equation}\label{eq:wilson_path2}
    f(\phi) = \prod_i h_1(W_{C_i}(\phi))
\end{equation}
where each $C_{i}$ is the elementary plaquette $\square$ for $i=1,...,L^2$,  $h_1(W_{C_i}(\phi))=1$ for $W_{C_i}(\phi)=\text{tr}(\mathbb{I})$ for the identity element $\mathbb{I} \in G$ and zero otherwise. This is because $A_v^g |\psi \rangle = |\psi \rangle$ for each $v, g$ from the above discussion and so is $A_v |\psi \rangle = |\psi \rangle$. And $B_f |\psi \rangle = |\psi \rangle$ holds due to the fact that for any finite group $\text{tr}(g)=\text{tr}(\mathbb{I})$ implies that $g=\mathbb{I}$.
}

\section{VI. Neural Network Architecture and Numerical Details}

In this section we provide the details for the architecture of the gauge equivariant neural network. There are two basic components for the network, which are the equivariant layer and the invariant layer. For the equivariant layer, we choose $\gamma_e$ to be $\sqcap$ for horizontal edge and  $\sqsubset$ for vertical edge and all $C_i^e$ to be $\square$ in Eq.~\ref{e:layer}. The function $h_e$ is taken to be convolutional neural network (CNN)and leaky relu activation function. One can further compose different equivariant layers in series or in parallel. For the invariant layer, it is expressed as $ f_e(\phi) = h_e\big(W_{C_1^e}(\phi),\ldots,W_{C_{n_e}^e}(\phi)\big)$, with all $C_i^e$ to be $\square$ and $h_e$ to be convolutional neural network. There are two outputs of $h_e$ in the invariant layer, which parameterizes the log amplitude and the phase of the wave function separately. The elu activation is used for the log amplitude output while the softsign activation is used for the phase output. The full network is made of composition of equivariant layers (Fig.~\ref{fig:architecture}(b)) followed by an invariant layer (Fig.~\ref{fig:architecture}(c)) in the end. For all the $8 \times 8, 10 \times 10, 12 \times 12$ experiments, we use neural network architecture as Fig.~\ref{fig:architecture} with kernel size of the CNN to be 4, 5, 5 respectively. For experiments in Fig.~\ref{fig:energy3}, the gauge equivariant neural network uses architecture in Fig.~\ref{fig:architecture}(a) with one channel of gauge equivariant blocks and the gauge invariant neural network uses architecture in Fig.~\ref{fig:architecture}(c), where the kernel size in all CNNs is 3. The RBM has hidden neurons three times as large as input neurons and outputs log amplitude and phase for the wave function.

Stochastic reconfiguration was performed with batch size 1000 and fixed learning rate 0.05. The number of iteration is 120 in general except that the experiments in $10 \times 10$ and $12 \times 12$ have iteration 150. The experiment in $12 \times 12$ starts with transfer learning of ansatzs optimized in $10 \times 10$. The convolution neural network parameters are initialized with orthogonal initialization \cite{saxe2014exact}. For sampling, we adopt the standard Monte Carlo sampling with single spin flip.

\section{VII. Observables and Quantities across Phase Transition}

\begin{figure}
        \includegraphics[width=1\linewidth]{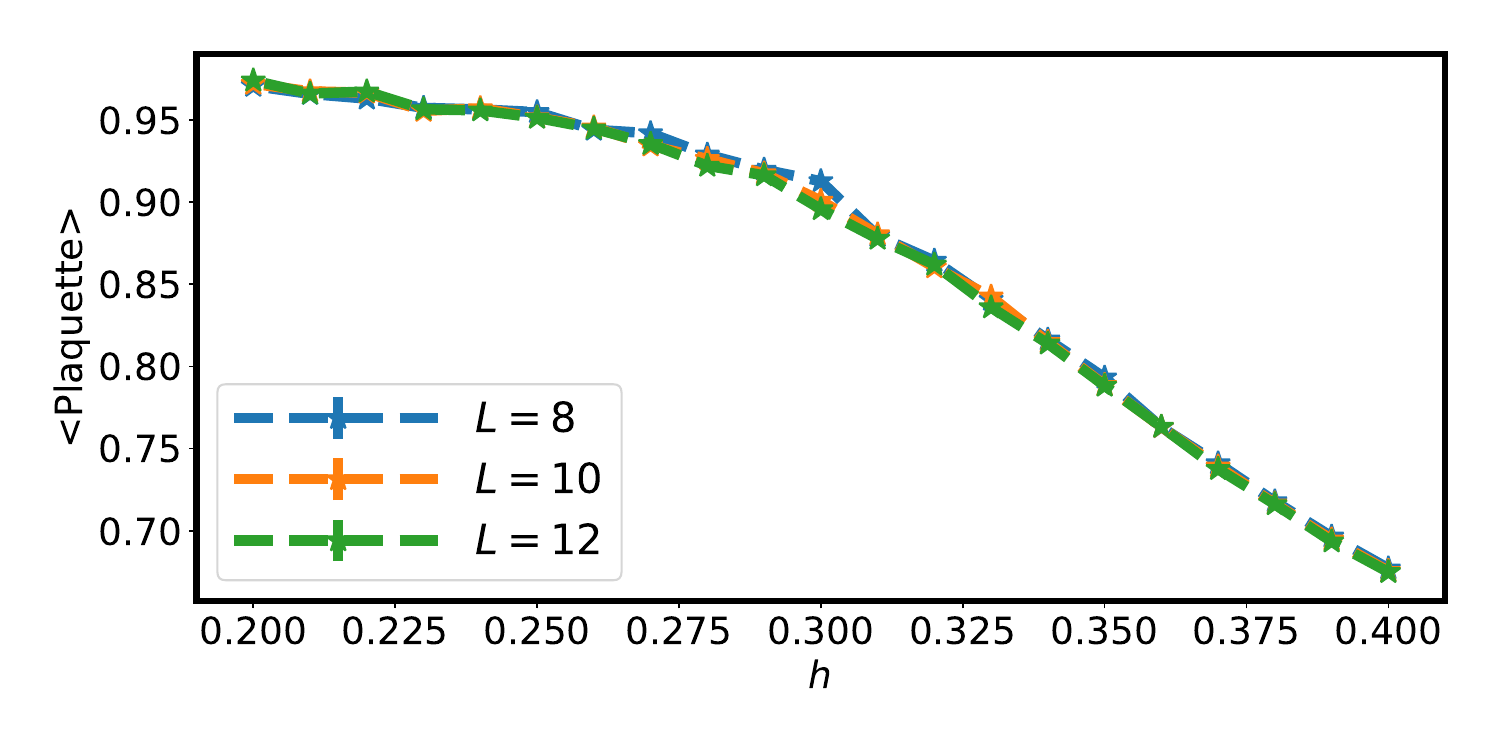}
   \caption{\label{fig:wilson_loop_avg} Average of expectation of plaquette $\langle \square \rangle$ over all plaquettes for different external field $h$ on $L \times L$ lattices. The change of the slope at around $h=0.3$ suggests a phase transition.}
\end{figure}

\begin{figure}
        \includegraphics[width=1\linewidth]{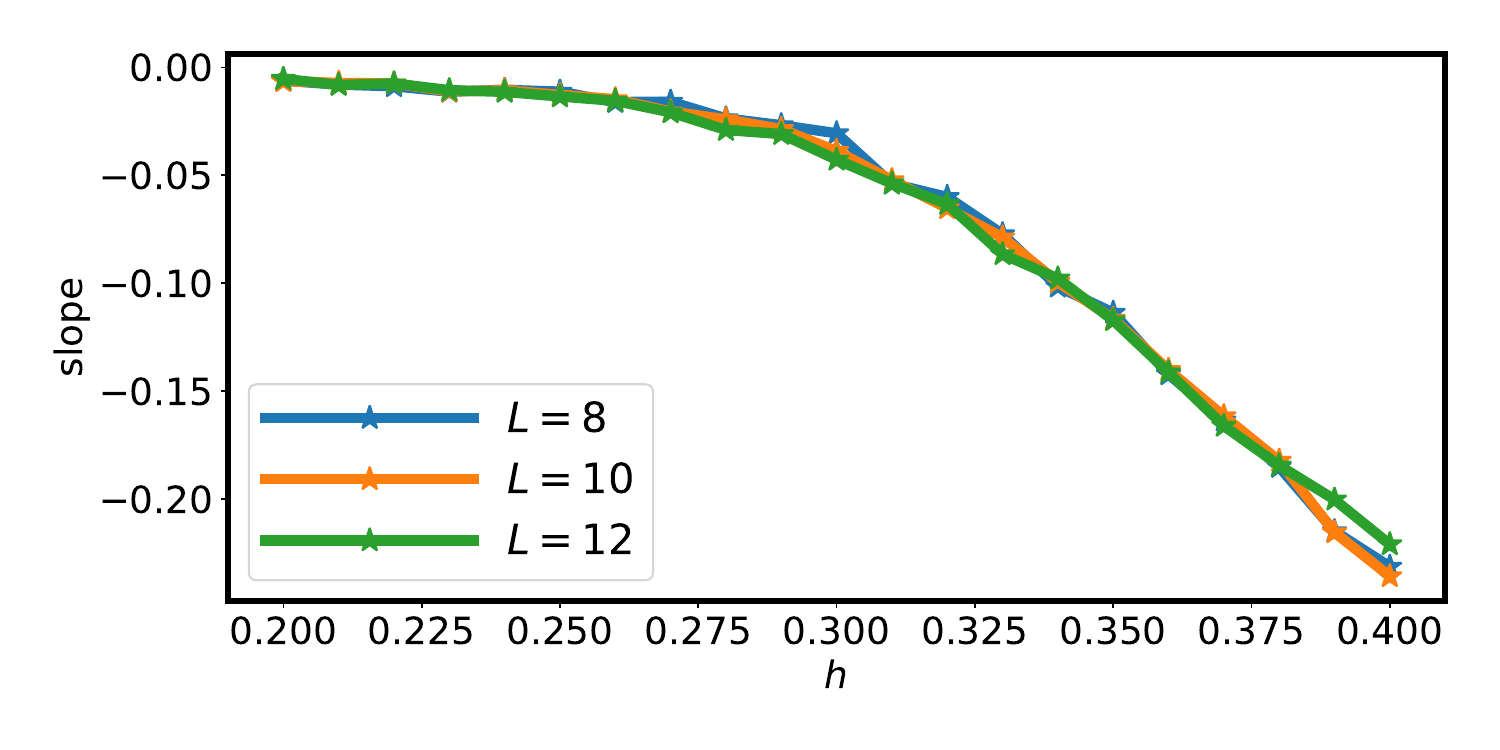}
   \caption{\label{fig:slope_area} Slope $k_1$ of the area ($A_C$) law fitting in log$\langle \hat{W}_C \rangle = k_1 A_C + b$ for different external field $h$ on $L \times L$ lattices. The changes of the slopes at around $h=0.3$ suggests a phase transition.}
\end{figure}

\begin{figure}
        \includegraphics[width=1\linewidth]{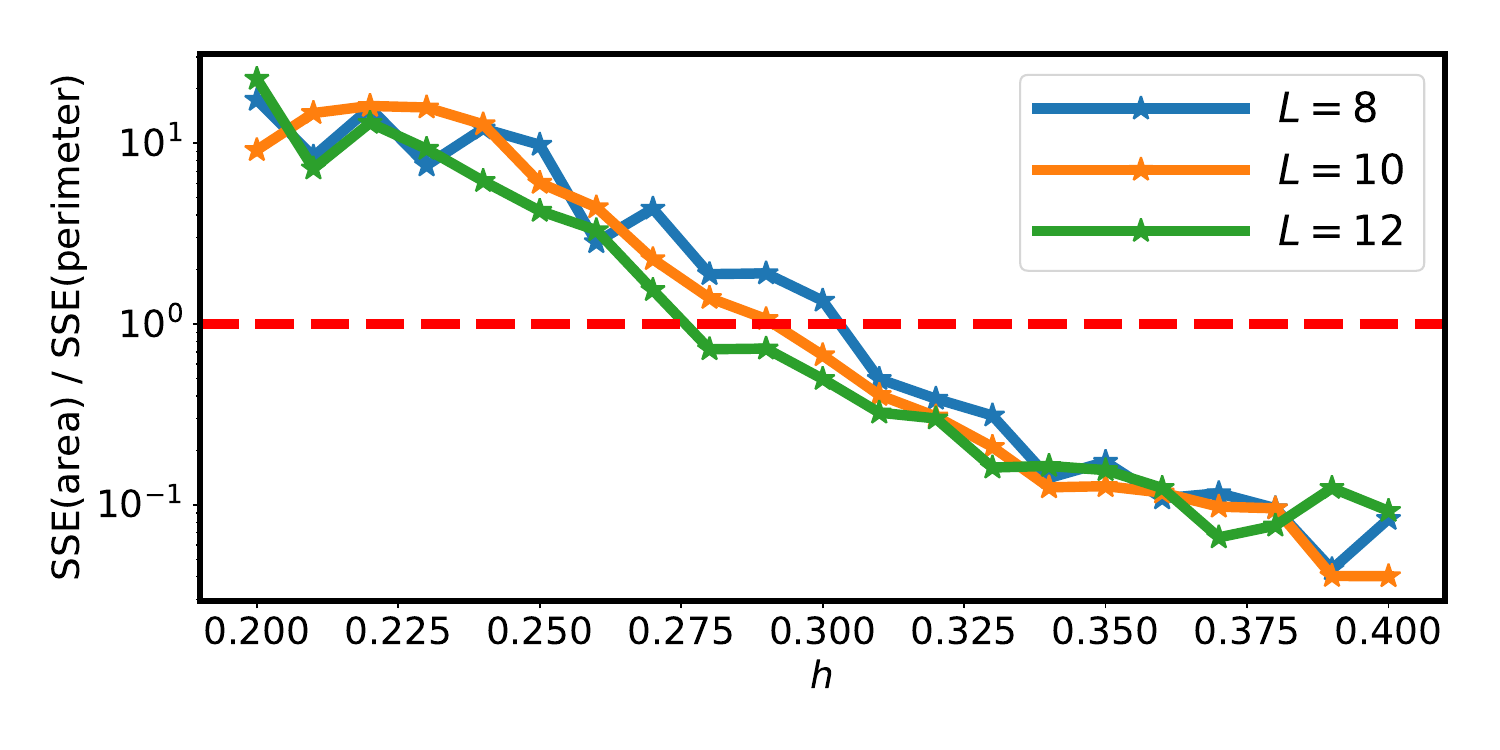}
   \caption{\label{fig:peri_area_slope} Ratio between the sum of squared estimate of errors (SSE) of the area law fitting  of $\log \langle \hat{W}_C \rangle$ and the SSE of the perimeter fitting of $\log \langle \hat{W}_C \rangle$ for different external field $h$ on $L \times L$ lattices. }
\end{figure}

\begin{figure}
        \includegraphics[width=1\linewidth]{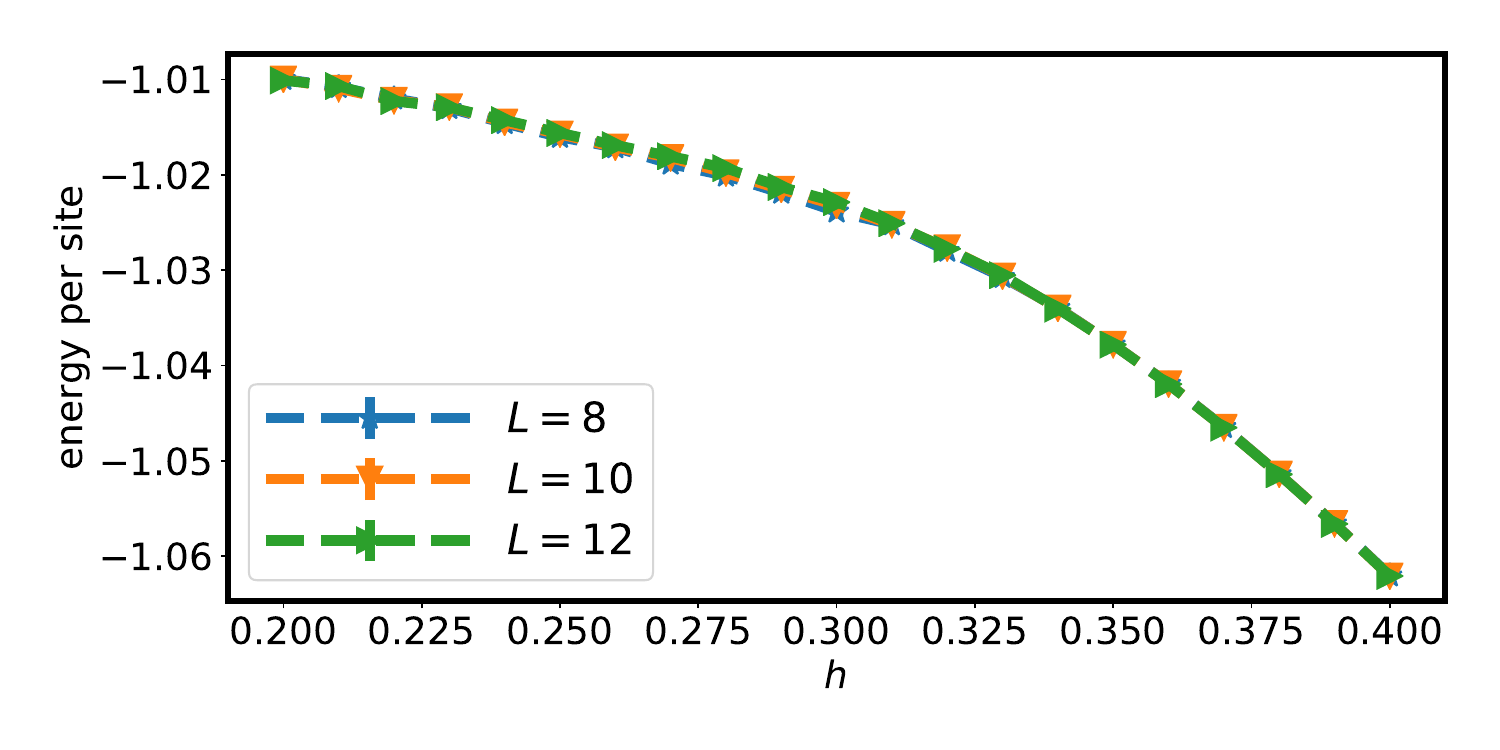}
   \caption{\label{fig:energy_all} Energy per site for different external field $h$ on $L \times L$ lattices.}
\end{figure}

\begin{figure}
        \includegraphics[width=1\linewidth]{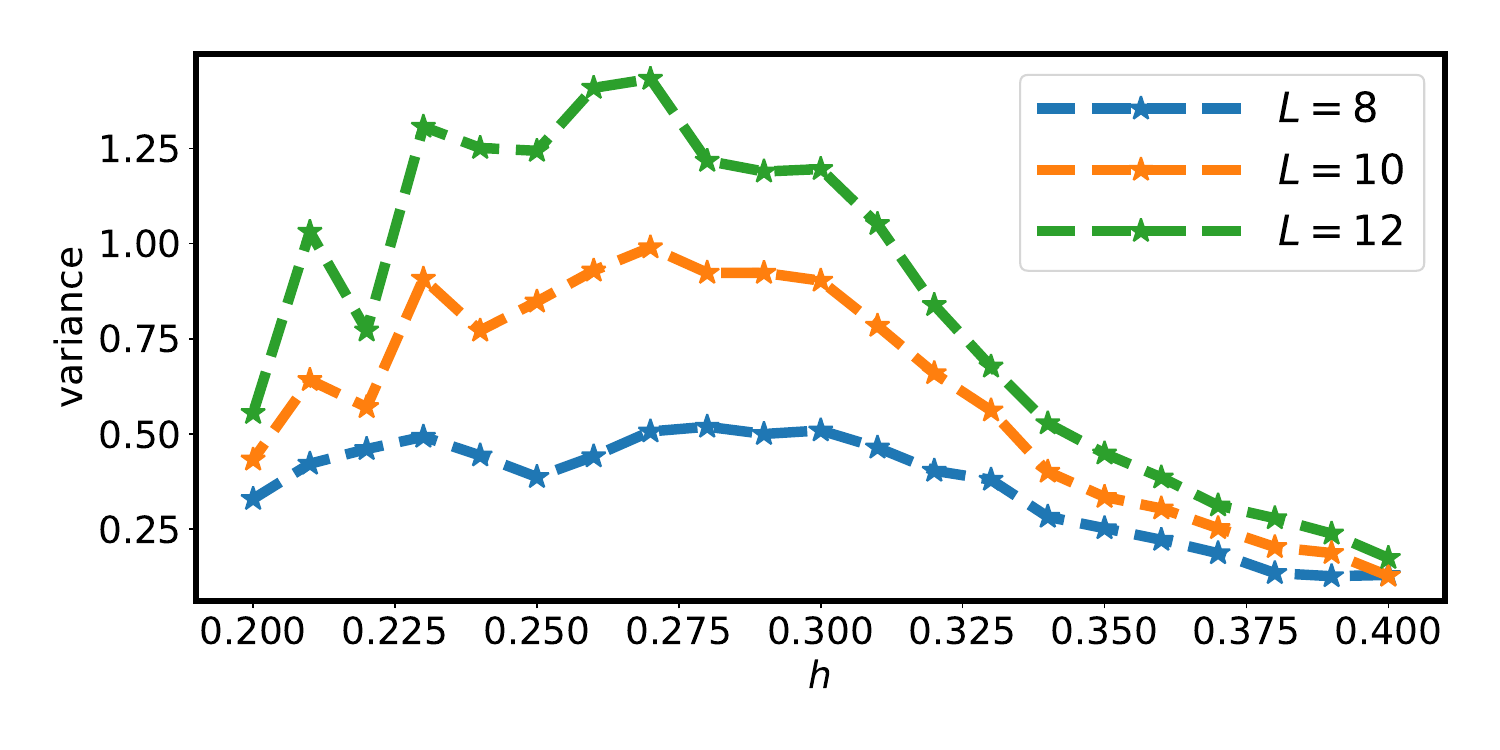}
   \caption{\label{fig:variance_all} Variance for different external field $h$ on $L \times L$ lattices.}
\end{figure}

We provide further figures (see Fig.~\ref{fig:wilson_loop_avg},~\ref{fig:slope_area},~\ref{fig:peri_area_slope},~\ref{fig:energy_all},~\ref{fig:variance_all}) on observables and quantities across the phase transition.

\end{document}